\documentclass[a4paper,fleqn,usenatbib]{mnras}

\usepackage{newtxtext,newtxmath}

\usepackage[T1]{fontenc}
\usepackage{ae,aecompl}

\usepackage{graphicx}	
\usepackage{amsmath}	
\usepackage{amssymb}	
\usepackage{enumitem}
\usepackage{lineno}

\newcommand{\lsim}{\protect\raisebox{-0.8ex}{$\:\stackrel{\textstyle <}{\sim}\:$}} 
\newcommand{\gsim}{\protect\raisebox{-0.8ex}{$\:\stackrel{\textstyle >}{\sim}\:$}}

\newcommand{\hMpc}{\ensuremath{h^{-1}\ {\rm Mpc}}}
\newcommand{\hMpck}{\ensuremath{h\ {\rm Mpc}^{-1}}}

\title[Iterative reconstruction]{An iterative reconstruction of cosmological initial density fields}

\author[R. Hada et al.]{
Ryuichiro Hada$^{1, 2, 3}$\thanks{E-mail: ryuichiro.hada@cfa.harvard.edu}
and Daniel J. Eisenstein$^{1}$
\\
$^{1}$Harvard-Smithsonian Center for Astrophysics, 60 Garden St., Cambridge, MA 02138, USA\\
$^{2}$Astronomical Institute, Tohoku University, Aoba-ku, Sendai 980-8578, Japan\\
$^{3}$Division for Interdisciplinary Advanced Research and Education, Tohoku University, Aoba-ku, Sendai 980-8578, Japan
}

\date{Accepted XXX. Received YYY; in original form ZZZ}

\pubyear{2017}

\begin{document}
\label{firstpage}
\pagerange{\pageref{firstpage}--\pageref{lastpage}}
	\maketitle

\begin{abstract} We present an iterative method to reconstruct the linear-theory initial conditions from the late-time cosmological matter density field, with the intent of improving the recovery of the cosmic distance scale from the baryon acoustic oscillations (BAOs). We present tests using the dark matter density field in both real and redshift space generated from an $N$-body simulation. In redshift space at $z = 0.5$, we find that the reconstructed displacement field using our iterative method are more than 80\% correlated with the true displacement field of the dark matter particles on scales $k < 0.10\hMpck$.  Furthermore, we show that the two-point correlation function of our reconstructed density field matches that of the initial density field substantially better, especially on small scales ($< 40\hMpc$).  Our redshift-space results are improved if we use an anisotropic smoothing so as to account for the reduced small-scale information along the line of sight in redshift space.
\end{abstract}
 
\begin{keywords}
large-scale structure of Universe -- dark matter -- distance scale -- cosmology:
theory
\end{keywords}

\section{Introduction} \label{sec:intro}

The baryon acoustic oscillation (BAO) plays an important role in the measurement of the cosmological distance scale and the determination of the cosmological parameters. With measurements of the cosmic microwave background (CMB), we can detect the acoustic signature as a harmonic sequence of oscillations in the temperature and polarization power spectra and obtain the highly precise cosmological information, such as about the baryon and dark matter densities and distance to the last-scattering surface \citep[for recent results, see][]{2016A&A...594A..13P}. In the context of the galaxy distribution on large scales, the acoustic signature is weaker but can still be detected as a peak in the correlation function \citep[][]{1970ApJ...162..815P,1970Ap&SS...7....3S,2005MNRAS.362..505C,2005ApJ...633..560E}. Fortunately, the scale of the peak is very large ($\sim 150$ Mpc), allowing the peak to be preserved even to the present time and to be well-modeled with low-order perturbation theory. The measurement of the cosmological distance scale using the BAO peak in the correlation function has been used as one of the strongest probes for investigating the property of dark energy \citep[for review, see][]{2013PhR...530...87W}, whose effects on the evolution of Universe dominates at late times \citep[for recent results, see][]{2017MNRAS.470.2617A}.

However, the nonlinear structure formation in the late-time blurs out the BAO peak in the correlation function and thus reduces the precision in the measurement of the distance scale \citep[][]{1999MNRAS.304..851M,2005Natur.435..629S,2005MNRAS.362L..25A,2005ApJ...633..575S,2006ApJ...651..619J,2007APh....26..351H,2008MNRAS.383..755A}. This degradation of the BAO peak is caused by the differential motion of pairs  of galaxies, mainly driven by gravitational potentials on large scales, e.g., bulk flows and supercluster formation \citep[][]{2007ApJ...664..660E}. \citet{2007ApJ...664..675E} showed that we can restore the BAO peak using a reconstruction method. The map of galaxies that are intended to measure the correlation function is also a map of the large-scale gravitational potential field. Therefore, from the same galaxy map, we can estimate the displacement of galaxies from the initial to the final location and move galaxies back to the original position. A particular version of this method, here called {\it standard reconstruction}, has been applied to galaxies survey and confirmed to reduce the distance error \citep[e.g., by a factor of 1.8 at $z=0.35$, ][]{2012MNRAS.427.2132P}.

Although standard reconstruction makes the distance measurement more precise, further improvements are desired for current and upcoming galaxy surveys, such as DESI\footnote{\url{http://desi.lbl.gov/}} \citep[][]{2016arXiv161100036D}, PFS\footnote{\url{http://pfs.ipmu.jp/}} \citep[][]{2014PASJ...66R...1T}, and Euclid\footnote{\url{https://www.euclid-ec.org/}} \citep[][]{2011arXiv1110.3193L}.  In fact, some types of reconstruction methods beyond the standard reconstruction have been proposed \citep[e.g.,][for details, see Section~\ref{sec:prev}]{2010ApJ...720.1650S,2012JCAP...10..006T,2017PhRvD..96b3505S,2017ApJ...841L..29W,2017ApJ...847..110Y,2017PhRvD..96l3502Z,2018PhRvD..97d3502Z,2018PhRvD..97b3505S}. 

In this paper, we present a new {\it iterative} reconstruction method. As in previous work, the idea is to repeat the calculation so as to improve the inference of the initial density field. Here we start with the iterative method motivated by \citet{1999MNRAS.308..763M} and develop a reconstruction method that converges in practical cases.  This is advantageous compared to past iterative methods in that we don't need to set the number of iteration, nor does the result depend on the numerical path by which we achieve convergence.  Further, our method provides a clean way to incorporate redshift-space distortions, without the spurious survey boundary artifacts that the standard reconstruction method creates.

This paper is organized as follows: we review the standard reconstruction method in Section~\ref{sec:sta_tec} and then develop an iterative reconstruction method in Section~\ref{sec:Itera}. In Section~\ref{sec:Resul}, we show the improvement of reconstruction compared with the standard method, discuss the effect of the anisotropy in redshift space, and then demonstrate the two-point correlation function. Finally, we conclude in Section~\ref{sec:Conc}.

\section{Standard Technique} \label{sec:sta_tec}

In this section, we review Lagrangian perturbation theory (LPT) and the original BAO reconstruction technique. The notation used in the following is based on that of \citet{1999MNRAS.308..763M}.

\subsection{\label{sec:Bas}Basics}

In LPT \citep[for review, see e.g.][]{2002PhR...367....1B}, the dynamics of objects are described by the displacement field ${\bf S}$ which maps the initial particle position ${\bf q}$ into the final Eulerian particle position ${\bf x}$:
\begin{eqnarray}
	{\bf x}({\bf q}, t)	= {\bf q} + {\bf S}({\bf q}, t).  \label{eq:x_q}
\end{eqnarray}
We can obtain a set of equations for the displacement ${\bf S}$ from the Euler-Poisson system of equations. The system is non-linear for ${\bf S}$ but can be solved perturbatively for small displacements. The linear solution ${\bf S}^{(1)}$ is  
\begin{eqnarray}
	\nabla \cdot {\bf S}^{(1)}({\bf q}) 
    &=&  - \delta_{\rm L}({\bf q}) =  - \frac{D(t)}{D(t_{i})}\delta(t_{i},{\bf q}),  \label{eq:ZA}
\end{eqnarray}
where $\delta_{\rm L}$ is the linear matter density contrast, $\delta(t_{i})$ is the initial density contrast, and $D$ is the linear growth factor. This relation corresponds to the Zel'dovich approximation \citep{1970A&A.....5...84Z}. Furthermore, the 2nd order solution ${\bf S}^{(2)}$ is written in terms of ${\bf S}^{(1)}$: 
\begin{eqnarray}
\nabla \cdot {\bf S}^{(2)}({\bf q}) 
    &=&  - \frac{3}{14} \Omega_{m}^{-1/143}\left(S^{(1)}_{a,a}S^{(1)}_{b,b} - S^{(1)}_{a,b}S^{(1)}_{b,a}\right),  \label{eq:2nd}
\end{eqnarray}
where $\Omega_{m}$ is the (time-dependent) matter density parameter and $S_{a, b} \equiv \partial S_a / \partial q_b$.

From the continuity equation, the density contrast can be described by the displacement ${\bf S}$:
\begin{eqnarray}
	\det \left[\delta^{\rm K}_{ab} + S_{a, b} \right] = \frac{\rho({\bf q})}{\rho({\bf x})} = \frac{1}{1 + \delta({\bf x})},  \label{eq:continuity}
\end{eqnarray}
where $\delta^{\rm K}_{ab}$ is the Kronecker delta and $\delta({\bf x}) = \rho({\bf x})/\bar{\rho}-1$ is the matter density contrast in terms of the Eulerian coordinate ${\bf x}$. In the second equality, we used the fact that the density contrast at the initial time can be neglected: $\rho({\bf q}) = \bar{\rho}$. Using the principal invariants of the 1st derivative tensor of ${\bf S}$, the determinant in the right hand side is 
\begin{eqnarray}
	\det \left[\delta^{\rm K}_{ab} + S_{a, b} \right] = 1 + \mu_{1}({\bf S}) + \mu_{2}({\bf S}) + \mu_{3}({\bf S}),  \label{eq:det_id}
\end{eqnarray}
where $\mu_{1}({\bf S}) = S_{a, a} = \nabla \cdot {\bf S}$, $\mu_{2}({\bf S}) = (S_{a, a}S_{b, b} - S_{a, b} S_{b, a})/2$, and $\mu_{3}({\bf S}) = \det(S_{a, b})$. Note that rotational fields would imply that $S_{a,b}\ne S_{b,a}$; however, the equation above is true regardless of whether or not ${\bf S}$ is irrotational.

The redshift-space ${\bf s}$ and the real-space ${\bf x}$ coordinates are related by
\begin{eqnarray}
	{\bf s}({\bf q}, t) = {\bf x}({\bf q}, t) + \frac{1}{aH} [{\bf v}({\bf q}, t) \cdot \hat{\bf z}]\hat{\bf z},
	\label{eq:s}
\end{eqnarray}
where $a(t)$ is the scale factor, $H(t)$ is the Hubble parameter, and $\hat{\bf z}$ is the unit vector pointing along the line of sight. ${\bf v}$ is the (physical) peculiar velocity at the position ${\bf q}$, which can be written in terms of ${\bf S}$ using Eq.~(\ref{eq:ZA}) and (\ref{eq:2nd}):
\begin{eqnarray}
	{\bf v}({\bf q}, t) = a \frac{{\rm d}{\bf S}}{{\rm d}t} = \frac{{\rm d} D}{{\rm d}t} \frac{a}{D} ({\bf S}^{(1)} + 2{\bf S}^{(2)}).  \label{eq:v}
\end{eqnarray}
Here we used the fact that the time dependence of ${\bf S}^{(2)}$ corresponds to $D^{2}(t)$. Then, we can introduce the redshift-space displacement ${\bf S}^{(s)}$ by
\begin{eqnarray}
	{\bf s}({\bf q}, t) = {\bf q} + {\bf S}^{(s)}({\bf q}, t),   \label{eq:s2}
\end{eqnarray}
where 
\begin{eqnarray}
	{\bf S}^{(s)}({\bf q}, t) = {\bf S}^{(1)} + {\bf S}^{(2)} + f[({\bf S}^{(1)} + 2{\bf S}^{(2)}) \cdot \hat{\bf z}]\hat{\bf z}.  \label{eq:S^s}
\end{eqnarray}
Here $f = {\rm d} \ln D/ {\rm d} \ln a$ is the linear growth rate. As noted by \citet{1999MNRAS.308..763M}, ${\bf S}^{(s)}$ can be rotational even if ${\bf S}$ is not.

Finally, we introduce the relation between the density fields in real and redshift spaces.    
In linear regime, we can relate the redshift-space galaxy density field $\tilde{\delta}^{g}_{s}$ in Fourier space to that in real space $\tilde{\delta}^{g}$ as follows \citep{1987MNRAS.227....1K}: 
\begin{eqnarray}
	\tilde{\delta}_{s}^{g}({\bf k}) = (1 + \beta \mu^{2})\tilde{\delta}^{g}({\bf k}).  \label{eq:kaiser}
\end{eqnarray}
Here $\mu = k_{z} / k$ ($z$ is the line-of-sight direction) and $\beta = f/b$, where $b$ is the linear galaxy bias, by which, assuming linear theory, the galaxy density field can be described by the matter density field: $\delta^{g} = b \delta$. 

\subsection{\label{sec:Sta}Standard reconstruction}

The original BAO reconstruction technique was introduced by \citet{2007ApJ...664..675E}.
Using Eq.~(\ref{eq:kaiser}), the procedure of standard reconstruction is summarized as follows \citep[e.g.][]{2011ApJ...734...94M, 2012MNRAS.427.2132P, 2016MNRAS.460.2453S}:
\begin{enumerate}
\setlength{\leftskip}{5mm}
\item \label{step_1} Using the final galaxy density field $\delta_{s}^{g}$ instead of the linear matter density field $\delta_{\rm L}$ in Eq.~(\ref{eq:ZA}), we can estimate the displacement field ${\bf S}^{(1)}$ in Fourier space,
\begin{eqnarray}
	\tilde{\bf S}^{(1)}_{\rm st}({\bf k}) = \frac{i {\bf k}}{k^2} \frac{\tilde{\delta}_{s}^{g}({\bf k})}{b(1 + \beta \mu^{2})} G(k),  \label{eq:ZA_st}
\end{eqnarray}
where $G(k)$ is a smoothing filtering traditionally given by  
\begin{eqnarray}
	G(k) = \exp[- 0.5 k^{2} \Sigma^{2}]. \label{eq:G}
\end{eqnarray}
Here $\Sigma$ is the smoothing scale.

\item \label{step_2} Displace the galaxies by $- {\bf S}^{(s)}_{\rm st} = - {\bf S}^{(1)}_{\rm st} - f({\bf S}^{(1)}_{\rm st} \cdot \hat{\bf z})\hat{\bf z}$ (see Eq.~(\ref{eq:S^s})) to form the displaced galaxy field $D$, and displace the random particles, that are uniformly distributed, by $- {\bf S}^{(1)}_{\rm st}$ to form the displaced random field $S$.

\item \label{step_3} Define the reconstructed correlation function $\xi_{\rm st}$ by the Landy-Szalay estimator \citep[][]{1993ApJ...412...64L}:
\begin{eqnarray}
	\xi_{\rm st} = \frac{DD - 2DS + SS}{SS}, \label{eq:rec_st}
\end{eqnarray}
where $DS$, etc. are the number of pairs at a given separation between various sets of points (i.e.,  between $D$ and $S$ in the case with $DS$).
\end{enumerate}
The Gaussian function defined as a smoothing filtering in Eq.~(\ref{eq:G}) is applied to eliminate high-density contrast corresponding to small-scale modes and the effects of shell crossing.  
Although this means that we lose the displacement information on smaller scale than the smoothing scale $\Sigma$, most of the degradation of the BAO peak from non-linear structure formation comes from large scales and can be estimated from the large-scale velocity field \citep[e.g.][]{2007ApJ...664..675E, 2012JCAP...04..013T}. 
The reason why a procedure using the smoothing function works well has been investigated theoretically in LPT by \citet{2009PhRvD..79f3523P}. 

The standard reconstruction introduced above is a simple technique for restoring the BAO peak; however, there are some problems to be improved:

\begin{enumerate}
\setlength{\leftskip}{5mm}

\item \label{prob_1} {\it Use of the final density field rather than initial density field when computing LPT.}

Lacking other information, the standard method uses the final galaxy density field instead of the linear density field in Eq.~(\ref{eq:ZA_st}) to estimate the displacement. We would prefer a method that relates the displacements to the initial field, so as to correctly include the physics of LPT.

\item \label{prob_2} {\it Limited to 1st order perturbation theory}

The standard method takes account of the 1st order (the Zel'dovich solution) only in LPT when estimating the displacement ${\bf S}$. It is known that the Zel'dovich solution describes exactly the nonlinear behavior of one-dimensional perturbation \citep[e.g.][]{Mukhanov200511}; however, higher-order perturbation theory could improve the dynamical modeling of realistic asymmetric density fields.

\item \label{prob_3} {\it Redshift-space distortion modeling}

As long as we consider only 1st order Eularian perturbation theory, the Kaiser formula described as Eq.~(\ref{eq:kaiser}) is exactly correct. However, it has been found out that there are some deviations from the Kaiser limit even at large scales due to ``large-scale" velocity dispersion  \citep[e.g.][]{1998MNRAS.296...10H, 2004PhRvD..70h3007S}, and the Kaiser formula doesn't fully capture the redshift-space distortions of the Zel'dovich approximation.   

\item \label{prob_4} {\it Unmatched motions of data and random particles}

In step~\ref{step_2} of the standard reconstruction procedure, the galaxies and the random particles in redshift space are displaced by different displacements in order to partially enforce the Kaiser approximation to redshift-space distortions.  However, this leads to a substantial anomaly, in that any sharp variation in the survey selection function (such as from any survey boundary) will be displaced by different amounts in the data and random set, leading to apparent order-unity density contrasts even in the limit of zero shot noise.  It also makes the boundary of the post-reconstruction survey ill-defined.  While these density anomalies do not appear to cause substantial problems in correlation studies to date, they reveal the ad hoc nature of the treatment of the redshift-space distortions.  Moreover, one could worry that these anomalies will increase in upcoming surveys such as DESI or PFS where the survey density is rapidly varying in the radial direction due to the redshifting of an emission line through the OH forest of sky emission.  We would prefer a method that moves the density field $\delta$ itself, such as in \citet[][]{2017JCAP...09..012O}, or equivalently moves the data and random particles together, so that the Lagrangian boundary of the survey is warping under the displacement field but not becoming separated between data and random particles \citep[see][for another possible way of dealing with this problem]{2015MNRAS.450.3822W}.
 
\end{enumerate}

\section{New iterative method} \label{sec:Itera}

\subsection{\label{sec:Theor}Theoretical framework}
	
In this paper, we reconstruct the displacement and the linear density field at the same time using the iterative method motivated by \citet{1999MNRAS.308..763M}. Therefore, we review their procedure before explaining our method in detail. Substituting the redshift-space displacement Eq.~(\ref{eq:s2}) into Eq.~(\ref{eq:continuity}) and combining Eq.~(\ref{eq:ZA}) with that, \citet{1999MNRAS.308..763M} obtained the following solution:
\begin{eqnarray}
	&&\delta_{\rm L}({\bf q}, t)
	= \left[\frac{\delta_{s}({\bf s})}{1 + \delta_{s}({\bf s})} \right.
	\nonumber \\
         && \left. \qquad \qquad - \mu_{1}({\bf S}^{(1)}) + \mu_{1}({\bf S}^{(s)}) + \mu_{2}({\bf S}^{(s)}) + \mu_{3}({\bf S}^{(s)}) \right], 
		\nonumber \\ 
	\label{eq:ite_ME}
\end{eqnarray}
where $\delta_{s}$ is the evolved density contrast in redshift space. 
Note that we can obtain the ${\bf q}$-space density $\delta_{s}({\bf s}) = \delta_{s}\left[{\bf s} ({\bf q})\right]$ by interpolating the ${\bf s}$-space density $\delta_{s}({\bf s})$ at the positions ${\bf s}$ corresponding to ${\bf q}$ through Eq.~(\ref{eq:s2}). The overview of their iterative method is summarized as follows: using the $n$th guess for the linear density, calculate the $(n+1)$th displacement from Eq.~(\ref{eq:ZA}) (and (\ref{eq:S^s})), estimate the $(n+1)$th guess from the linear density from Eq.~(\ref{eq:ite_ME}), and then iterate these steps. This means that the ${\bf q}$-space density $\delta_{s}\left({\bf s} ({\bf q})\right)$ changes at each iteration. In addition, before starting the iteration, the ${\bf s}$-space density $\delta_{s}({\bf s})$ must be smoothed in advance to avoid the denominator of the first term in the right hand side of Eq.~(\ref{eq:ite_ME}) from approaching zero. However, this smoothing, which is different from the smoothing discussed in Section~\ref{sec:Sta}, decreases the small-scale fluctuations of the density field, smoothing the BAO feature in the correlation function contrary to our goal. 

In order to avoid this problem, we introduce some adjustments.  We assume that the linear density contrast can be separated into the large-scale part $\delta_{l}$ and residual part $\delta_{\rm res}$:
\begin{eqnarray}
	\delta_{\rm L}({\bf q},t) = \delta_{l}({\bf q},t) + \delta_{\rm res}({\bf q}).
	\label{eq:separation}
\end{eqnarray}
We consider a model in which only the large-scale portion creates displacements, which then advect the small-scale residual as a passive tracer. 
While this is obviously not correct on small scales, by using a smoothing filter to do the scale separation, we create a smooth transition between the two regimes.  In particular, we define the large scale field via
\begin{eqnarray}
	\tilde{\bf S}^{(1)}_{l}({\bf k}) &=& \frac{i {\bf k}}{k^2} \tilde{\delta}_{\rm L}({\bf k}) G(k),  \label{eq:shift_large} \\
	\delta_{l}({\bf q},t) &=& - \nabla \cdot {\bf S}_{l}^{(1)}({\bf q},t) = - \mu_{1}({\bf S}_{l}^{(1)}). 
	\label{eq:large}
\end{eqnarray}
Here we opt to use only first-order displacements, ${\bf S} = {\bf S}^{(1)}$, for simplicity and because we expect that the sparseness of realistic galaxy samples will require large enough smoothing scales that  first-order will be sufficient.

In other words, taking account of the fact that the degradation of the BAO peak is mainly caused by the large-scale velocity field, we neglect the gravitational effects of modes smaller than the smoothing scale $\Sigma$. This means that the residual part is assumed to have existed at the initial time: $\rho({\bf q}) = \bar{\rho}(1 +  \delta_{\rm res}({\bf q}))$.  Therefore, we need to modify the continuity equation as follows to make our scenario self consistent:
\begin{eqnarray}
	\det \left[\delta^{\rm K}_{ab} + S^{(s)}_{l|a, b} \right] = \frac{1 + \delta_{\rm res}({\bf q})}{1 + \delta_{s}({\bf s})}.  \label{eq:continuity_mo}
\end{eqnarray}
where ${\bf S}_{l}^{(s)}$ is related to ${\bf S}_{l}^{(1)}$ through Eq.~(\ref{eq:S^s}). Here we note that the assumption of the residual part existing at the initial time does not mean that there is no fluctuations on large scales at the later time. The numerator and denominator of the right hand side in Eq.~(\ref{eq:continuity_mo}) correspond to the inital time and the observed time, respectively. Therefore, we are rather keeping all modes at thier late-time amplitude through Eq.~(\ref{eq:continuity_mo}) and only apply gravity to estimate the displacement filed.    

Since the residual part $\delta_{\rm res}$ gets close to $0$ as $G(k) \to 1$, Eq.~(\ref{eq:continuity_mo}) becomes the original continuity equation if we don't consider the smoothing ($\Sigma \to 0$) and Eq.~(\ref{eq:separation}) leads to $\delta_{\rm L}({\bf q}) = \delta_{l}({\bf q})$ as expected, which means that the density contrast on all scales is displaced. On the other hand, at the opposite limit $G(k) \to 0\ (\Sigma \to \infty)$, the large-scale displacement field gets close to $0$, which corresponds to $\delta_{\rm L}({\bf q}) = \delta_{\rm res}({\bf q})$. Then it follows from Eq.~(\ref{eq:continuity_mo}) that $\delta_{\rm res}({\bf q}) = \delta_{\rm s}({\bf s})$ and therefore we find that at this limit, the observed density field is not reconstructed at all. In our method, we use a finite smoothing scale, between these limits, to avoid the small-scale complexities discussed above.  

Finally, redshift-space distortions coming from small-scale thermal 
motions, i.e., the Fingers of God effect, can cause small-scale density fluctuations to appear at larger scales along the line of sight.  These modes no longer accurately reflect large-scale gravitational dynamics.
To mitigate this, we seek to down-weight these density fluctuations in the smoothed density field and therefore introduce the parameter, $\mathcal{C}_{\rm ani}$, as follows: 
\begin{eqnarray}
	\mathcal{C}_{\rm ani} &\equiv& \Sigma_{\parallel}/\Sigma_{\perp}, \label{eq:C_ani}
\end{eqnarray}
where $\Sigma_{\parallel}$ and $\Sigma_{\perp}$ are the smoothing scales along the line of sight and the perpendicular directions, respectively:
\begin{eqnarray}
	G_{\rm ani}(k) &=& \exp[- 0.5 (k_{\perp}^{2}\Sigma^{2}_{\perp} + k_{\parallel}^{2}\Sigma^{2}_{\parallel} )]
    			   \nonumber \\
    			   &=& \exp[- 0.5 (k_{\perp}^{2} + k_{\parallel}^{2}\mathcal{C}_{\rm ani}^{2})\Sigma^{2}_{\perp}]. \label{eq:G_ani}
\end{eqnarray}
Effectively, this means that we use less of the line-of-sight density fluctuations in deriving the large-scale displacements \citep[see also][for anisotropic smoothing for reconstruction]{2016MNRAS.457.2068C}.

\subsection{\label{sec:imple}Implementation}

Our final goal is to find the linear density field $\delta_{\rm L}({\bf q})$ that solves Eq.~(\ref{eq:continuity_mo}) given the observed density $\delta_{s}({\bf s})$ and the definitions in Eq.~(\ref{eq:separation})-(\ref{eq:large}) and (\ref{eq:S^s}). To do so, we derive the solution in our context corresponding to Eq.~(\ref{eq:ite_ME}). Combining Eq.~(\ref{eq:separation}), (\ref{eq:large}) and (\ref{eq:continuity_mo}), we obtain
\begin{eqnarray}
	\frac{1 + \delta_{\rm L}({\bf q},t) + \mu_{1}({\bf S}_{l}^{(1)})}{1 + \delta_{s}({\bf s})} = \det \left[\delta^{\rm K}_{ab} + S^{(s)}_{l|a, b} \right].  
	\label{eq:continuity_mo_2}
\end{eqnarray}
Then using the expression of the determinant Eq.~(\ref{eq:det_id}), the solution for $\delta_{\rm L}$ is
\begin{eqnarray}
	&&\delta_{\rm L}({\bf q}, t) = - \mu_{1}({\bf S}_{l}^{(1)}) -1
		\nonumber \\ 
		&& \quad + (1 + \delta_{s}({\bf s}))\left[1 + \mu_{1}({\bf S}_{l}^{(s)}) + \mu_{2}({\bf S}_{l}^{(s)}) + \mu_{3}({\bf S}_{l}^{(s)}) \right].
		\nonumber \\ 
	\label{eq:ite}
\end{eqnarray}
In detail, we implement the modified iterative method as follows:
\begin{enumerate}
\setlength{\leftskip}{5mm}
\item \label{item:1} Distribute the galaxy particles on a grid using Triangular Shaped Cloud (TSC) scheme\footnote{There are some methods to assign particles to grid cells: Nearest Grid Point (NGP; "point"), Cloud In Cell (CIC; "linear"), and Triangular Shape Cloud (TSC; "quadratic")  \citep[see][for more details]{1988csup.book.....H}. In this paper, we employ TSC method achieving higher resolution in order to evaluate the performance of our iterative reconstruction method although it is computationally more expensive.} and calculate the observed density field in redshift space $\delta_{s}({\bf s})$ at each grid cell.

\item \label{item:2} Set the first guess for the linear density to $\delta_{\rm L}({\bf q}) = \delta_{s}({\bf s})$ with ${\bf S}_{l}^{(s)} = 0$.

\item \label{item:3} Estimate the displacements ${\bf S}_{l}^{(1)}$ and ${\bf S}_{l}^{(s)}$ using Eq.~(\ref{eq:shift_large}) and (\ref{eq:S^s}), respectively.

\item \label{item:4} Update the guess for the linear density $\delta_{\rm L}$ using Eq.~(\ref{eq:ite}).

\item \label{item:5} Repeat steps \ref{item:3} and \ref{item:4} until the linear density field converges.
\end{enumerate}

In principle, we seek a field $\delta_{\rm L}({\bf q})$ that provides a consistent solution to our equations.  In practice, there might be multiple solutions or troublesome domains of convergence for the above algorithm.  We therefore adopt two techniques to aid the iterative solver.
First, we treat the smoothing scale $\Sigma$ as an annealing parameter.
We start with a large value and reduce it gradually at each iteration until it reaches the effective smoothing scale, $\Sigma_{\rm eff}$, that is supposed to be applied when estimating the displacement we finally obtain \citep[e.g.][]{2017PhRvD..96b3505S}:
\begin{eqnarray}
	\Sigma_{\perp,n}  = \mbox{max}\left(\frac{\Sigma_{\rm ini}}{\mathcal{D}^{n}},\  \Sigma_{\rm eff}\right), \label{red_sm}
\end{eqnarray}
where $\Sigma_{\perp,n}$ is the smoothing scales along the perpendicular direction of the $n$th iteration and $\Sigma_{\rm ini}$ is the initial smoothing scale. $\mathcal{D}$ is a constant which is larger than 1. In the followings, we set $\Sigma_{\rm ini} = 20\hMpc$ and $\mathcal{D} = 1.2$.
This annealing is powerful because the largest-scale displacements are well predicted by linear theory and thereby can be recovered even from the observed final density field, yet applying these motions causes the small-scale features in the initial and final fields to come into phase with each other.  We stress that the parameters in the annealing have no physical meaning and the converged results should be independent of small changes in the exact annealing steps.

Second, following techniques used by \citet{1999MNRAS.308..763M}, we take our next value of the linear density field to be a weighted average of the computed value and the previous value.  Specifically, we use
\begin{eqnarray}
	\delta_{\rm L}^{(n)} = w \delta_{\rm L[ori]}^{(n)} + (1-w)\delta_{\rm L}^{(n-1)},  \label{eq:weight}
\end{eqnarray}
where $\delta_{\rm L}^{(n)}$ is the $n$th guess for the linear density, $\delta_{\rm L[ori]}^{(n)}$ is the left hand side of Eq.~(\ref{eq:ite}) at $n$th iteration, and the weight $w$ is within the range of $0 < w <1$. 
This is a standard gain in control theory, intended to damp out oscillations in the iteration.

Furthermore, we need a criterion to make sure whether the solution converges or not, and therefore define the ratio showing how the guess of the linear density changes compared with the last time:
\begin{eqnarray}
	r_{\rm con} \equiv \frac{\sum [\delta_{\rm L[ori]}^{(n)} - \delta_{\rm L}^{(n-1)}]^{2} }{\sum \delta_{s}^{2}},  \label{eq:con}
\end{eqnarray}
where $\sum$ means the summation over all grid cells. In this work, we repeat our procedure until $r_{\rm con}$ gets smaller than $0.01$.

\subsection{\label{sec:discu}Discussion}

Here we consider the advantages and limitations of our iterative method as well as the relation with problem \ref{prob_1}-\ref{prob_4} mentioned in Section~\ref{sec:Sta}.

First of all, the primary assumption of this method is the separation of the linear density field $\delta_{\rm L}$ into the large-scale part that is gravitationally active and the residual part that is simply advected (see Eq.~(\ref{eq:separation})). Although this is not physically exact, we can expect that this picture would describe properly the velocity field on much larger scales than the smoothing scale.  Like the standard reconstruction, one is advecting the unsmoothed final density field back to the initial field, keeping all of the density information in the one scalar field.  If the smoothing scale is large, then the method gracefully degrades back to no alteration.

The method is explicit about finding an initial density field that will evolve into the observed positions, on scales larger than the smoothing.  In this way, it redresses problem \ref{prob_1}: we no longer use the final density field as the source to find the Zel'dovich approximation displacements.  

As we avoid dividing by $1+\delta_s({\bf s})$, the results are insensitive to the grid cell size and finite particle sampling.  One applies the smoothing only when computing the displacement field and can use as fine a grid as one wants to track the density field information.  By keeping the density field on a grid, one can gain a computational convenience by immediately applying grid-based clustering statistics without returning to the particle set.

The method could be immediately altered to include the second-order Lagrangian perturbation theory evolution of the proposed $\delta_{\rm L}({\bf q})$, including the second-order redshift-space corrections, so as to address problem \ref{prob_2}.  We have opted to use only first order in this presentation, because we expect that applications to wide-field surveys will need to contend with galaxy bias and shot noise.  Getting more accurate answers in realistic applications will require more than simply upgrading from first to second-order in the matter perturbation theory.  We stress that although the Zel'dovich approximation is first order, it is known to be much more accurate than first-order Eulerian perturbation theory \citep[e.g.][]{1993MNRAS.260..765C, 2012JCAP...04..013T}, exactly because it is well poised to follow the bulk flows that dominate the degradation of the acoustic peak.

As we are starting from the Lagrangian evolution of an initial field, the results seamlessly include the large-scale redshift-space distortions (problem \ref{prob_3}).  Further, we are free to use an anisotropic smoothing, $\mathcal{C}_{\rm ani}$ (Eq.~(\ref{eq:C_ani})), which can help to avoid some impact from small-scale distortions, as we will see in Section~\ref{sec:ani_RSD}.

Finally, we have fully avoided problem \ref{prob_4}.  Once we have estimated the final density field $\delta_s$ from the comparison of data points to random points, we never return to the data and random points separately.  Redshift-space distortions have been included by their impact on the displacement field and the resulting determinant of the shear tensor, rather than some differential movement of the data points.

\subsection{\label{sec:prev}Previous work}

A series of reconstruction methods beyond standard reconstruction has been proposed so far  \citep[see, for an overview,][]{2017PhRvD..96b3505S}. These procedures are generally composed of two parts: 1) estimating the displacement from the initial to the final position, and 2) computing the linear density field from the displacement. 

In particularly, some methods using ``iterative" procedures in the first step have been proposed. These methods have main differences in the smoothing filtering ({\it pattern A}): with a fixed smoothing scale \citep{2010ApJ...720.1650S}, optimal filtering (Wiener filtering) \citep{2012JCAP...10..006T}, and reducing the smoothing scale as the iteration progresses \citep{2017PhRvD..96b3505S}. 

On the other hand, another type of iterative reconstruction methods have been proposed. These methods are based on numerical solutions to the nonlinear partial differential equation that governs the mapping between the initial particle position ${\bf q}$ and the final Eulerian particle position ${\bf x}$ ({\it pattern B}). In order to solve the differential equation for the matter density field in real space, \citet{2017PhRvD..96l3502Z} used a moving mesh approach, which was subsequently applied to the BAO measurement~\citep{2017ApJ...841L..29W}, halo fields~\citep{2017ApJ...847..110Y}, and redshift-space distortions~\citep{2018PhRvD..97d3502Z}. Recently, \citet{2018PhRvD..97b3505S} proposed a multigrid relaxation method to solve the differential equation.

Although it is found that these iterative methods achieve substantial improvements compared to the standard reconstruction, it is hard to compare the performances of them quantitatively. The main qualitative difference between our method and other iterative methods is that we take account of only the 1st order in LPT (simpler than {\it pattern B}) while making the solution converge by using Eq.~(\ref{eq:ite}) in the iteration process (more reliable than {\it pattern A}).

\begin{figure*}
\begin{tabular}{cc}
 \begin{minipage}{0.5\hsize}
  \begin{center}
   \includegraphics[width=90mm]{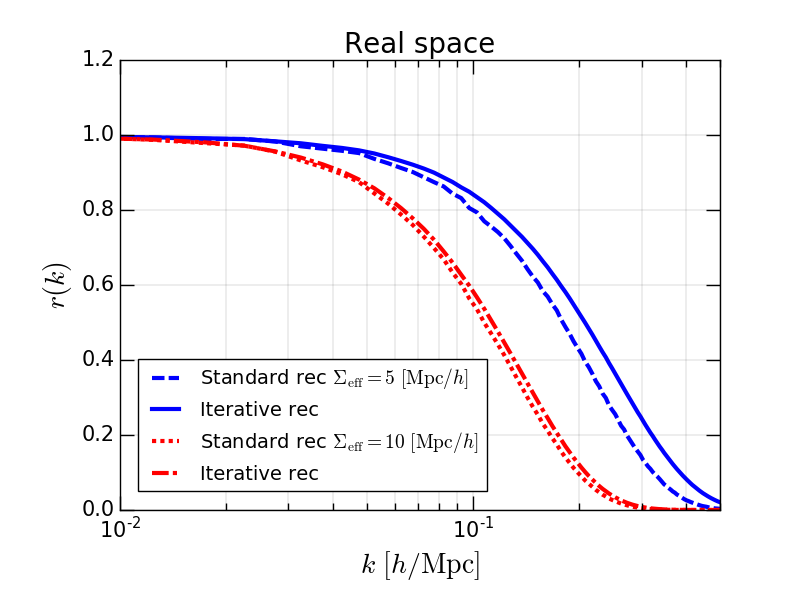}
  \end{center}
 \end{minipage}
 \begin{minipage}{0.5\hsize}
  \begin{center}
   \includegraphics[width=90mm]{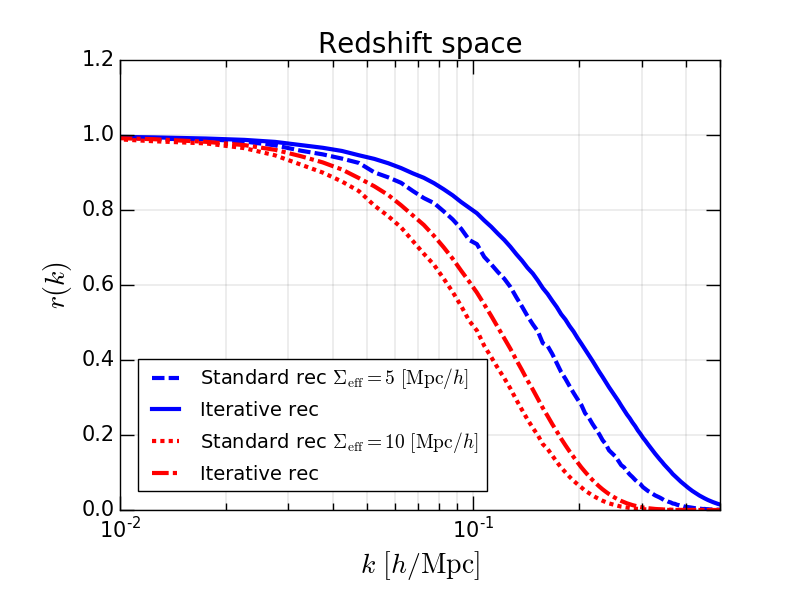}
  \end{center}
\end{minipage}
\end{tabular}
  \caption{\label{x-corr_r} Cross-correlation coefficient with the true displacement field of the dark matter particles, $r(k)$, in real space (left panel) and redshift space (right panel). The solid (in blue) and dot-dashed (in red) lines show the coefficients at $z=0.5$ for our iterative reconstruction method with 5$h^{-1}$ and 10$\hMpc$ smoothing scale, respectively. The dashed (in blue) and dotted (in red) lines show the one for the standard reconstruction method with 5$h^{-1}$ and 10$\hMpc$ smoothing scale, respectively.}
\end{figure*}

\begin{figure*}
\begin{tabular}{cc}
 \begin{minipage}{0.5\hsize}
  \begin{center}
   \includegraphics[width=75mm]{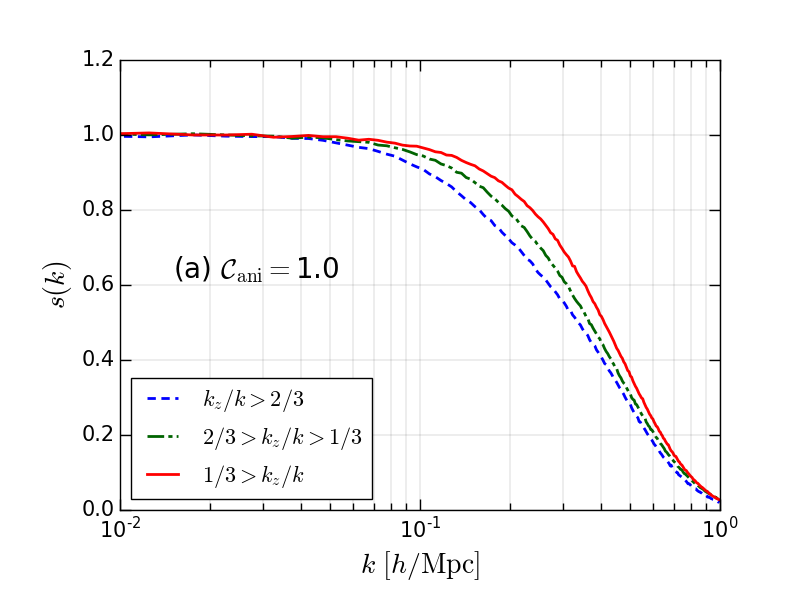}
  \end{center}
\end{minipage} 
 \begin{minipage}{0.5\hsize}
  \begin{center}
   \includegraphics[width=75mm]{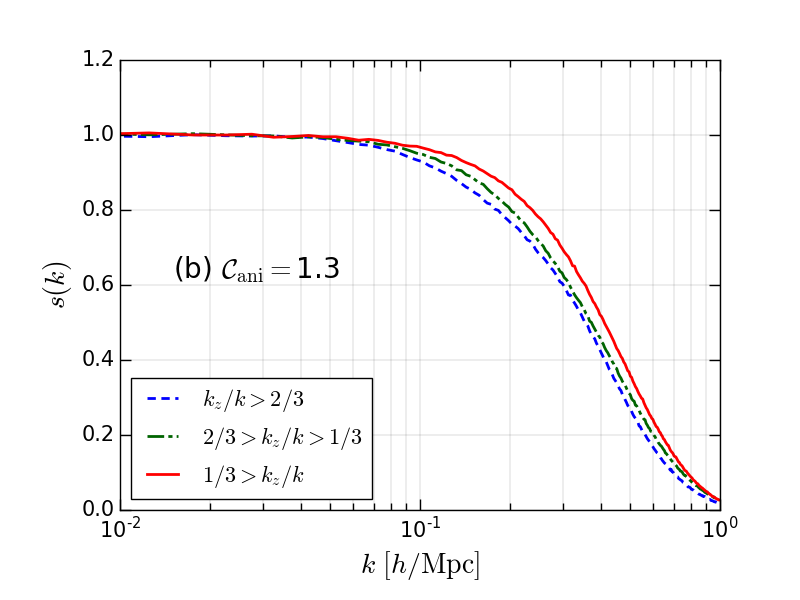}
  \end{center}
\end{minipage} \\
 \begin{minipage}{0.5\hsize}
  \begin{center}
   \includegraphics[width=75mm]{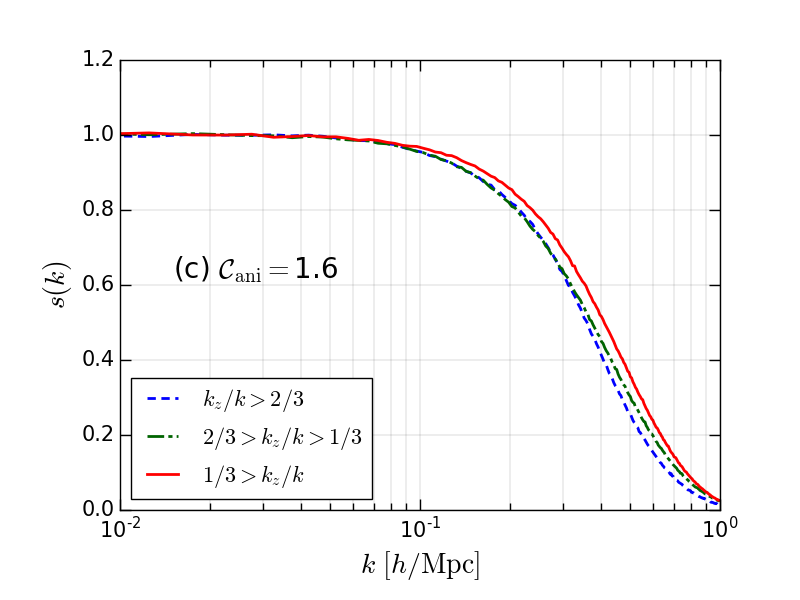}
  \end{center}
\end{minipage} 
 \begin{minipage}{0.5\hsize}
  \begin{center}
   \includegraphics[width=75mm]{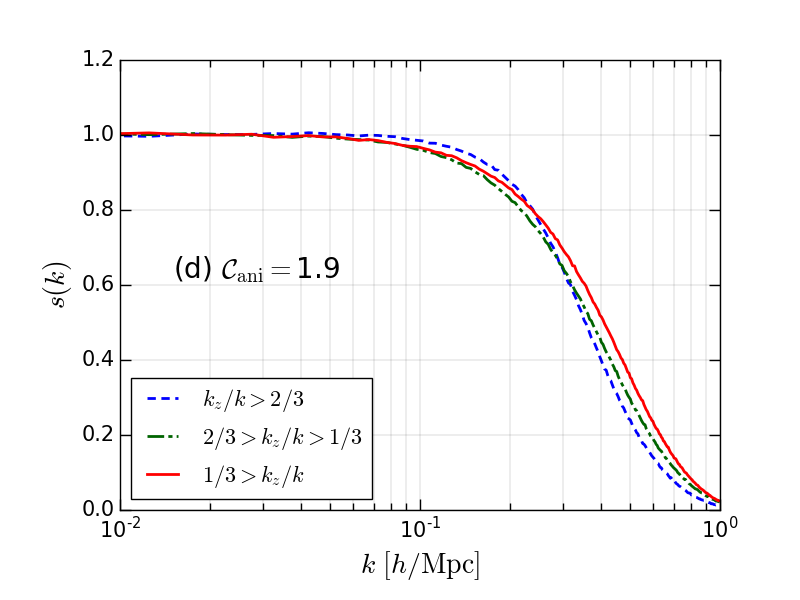}
  \end{center}
\end{minipage}
\end{tabular}
  \caption{\label{x_corr_s} Cross-correlation coefficient with the initial density field, $s(k)$, in redshift space for various $\mathcal{C}_{\rm ani}$. The solid (in red), dot-dashed (in green), and dashed (in blue) lines show the coefficients for $5\hMpc$ smoothing scale at $z=0.5$ averaged over the  directions: $1/3 > k_{z}/k$, $2/3 > k_{z}/k > 1/3$, and $k_{z}/k > 2/3$, respectively.}
\end{figure*}

\section{Results} \label{sec:Resul}

\subsection{\label{sec:nbody}{\it N}-body simulation and parameter setting}

To evaluate the performance of our iterative reconstruction method, we apply that to the matter density field produced with {\sc abacus}, which is an extremely fast and accurate $N$-body code for cosmological simulations \citep[][Ferrer et al., in preparation; Metchnik \& Pinto, in preparation]{2016MNRAS.461.4125G, 2018ApJS..236...43G}. $3200^{3}$ particles are simulated in a box of length $L = 1600 \hMpc$ with the parameters based on \citet{2016A&A...594A..13P}. We use, in particular, $640^{3}$ particles ($\sim 1 \%$) chosen randomly from  one realization at redshift $z = 0.5$. Moreover, we perform the reconstruction of the matter density and the calculation of some types of correlations using a $640^{3}$ grid according to the number of particles. In this work, we try two types of the effective smoothing scales: $\Sigma_{\rm eff} = 5 h^{-1}$ and $10\hMpc$, and set the weight $w = 0.3$ and $0.5$, respectively, so that the estimation of linear density field doesn't diverge. Note that in this setting, the number of iteration that is required for the convergence condition is $17$ times and $9$ times, respectively. We emphasize that the weight and the number of iteration are likely application-specific. 

\subsection{\label{sec:x-corr_shift}Cross correlation for the displacement}

To show how our iterative reconstruction method improves the standard reconstruction method, we focus on the cross-correlation between the reconstructed displacement field and the true displacement field of the dark matter particles which is traced in $N$-body simulation. For this purpose, we introduce the correlation coefficient for the displacement as follows:
\begin{eqnarray}
r(k) \equiv \frac{\langle \tilde{{\bf S}}_{l}^{(s)} \cdot \tilde{{\bf S}}^{*}_{\rm tru} \rangle}{\langle |\tilde{{\bf S}}_{\rm tru} |^2 \rangle}, \label{eq:coe_r}
\end{eqnarray}
where $\tilde{{\bf S}}_{l}^{(s)}$ is the displacement in Fourier space that is reconstructed with our iterative method and $\tilde{{\bf S}}_{\rm tru}$ is the true displacement in Fourier space along which the dark matter particles have moved from the initial to the final position. 

Fig.~\ref{x-corr_r} shows the coefficient $r(k)$ as a function of the wavenumber $k$ in real space (left panel) and redshift space (right panel). In real space, both of displacements reconstructed using the standard method and our method, with $10\hMpc$ smoothing scale, are more than 80\% correlated with the true displacement on scales $k \lsim 0.06 \hMpck$ and we can see that our method is correlated slightly well than the standard method. On the other hand, with $5\hMpc$ smoothing scale, the displacements reconstructed using the standard method and our method are more than 80\% correlated on scales $k \lsim 0.10h$ and $0.12 \hMpck$, respectively. This means that using our method, the reconstruction in real space is definitely improved compared with the standard method. Moreover, we see that the case with $5\hMpc$ smoothing scale is more correlated than the case with $10\hMpc$ because the smaller the smoothing scale gets, the smaller the scale that we are supposed to take account of through Eq.~(\ref{eq:shift_large}) each iteration becomes.

In redshift space, compared with real space, we can see that the cases based on the standard reconstruction are less correlated and that the difference between the both methods gets larger. This suggests that our method can take account of the effect of the redshift-space distortion well. However, at the same time, the correlation coefficient for our method in redshift space becomes smaller slightly than the one in real space, which means that the anisotropy in redshift space has an effect on the process of our iterative method somewhat.

\begin{figure*}
\begin{tabular}{ccc}
 \begin{minipage}{0.33\hsize}
  \begin{center}
   \includegraphics[width=65mm]{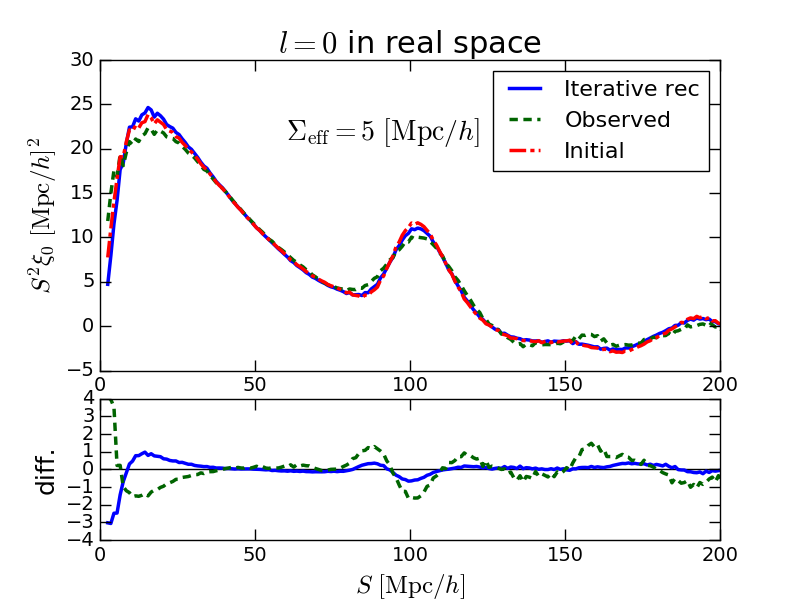}
  \end{center}
\end{minipage} 
 \begin{minipage}{0.33\hsize}
  \begin{center}
   \includegraphics[width=65mm]{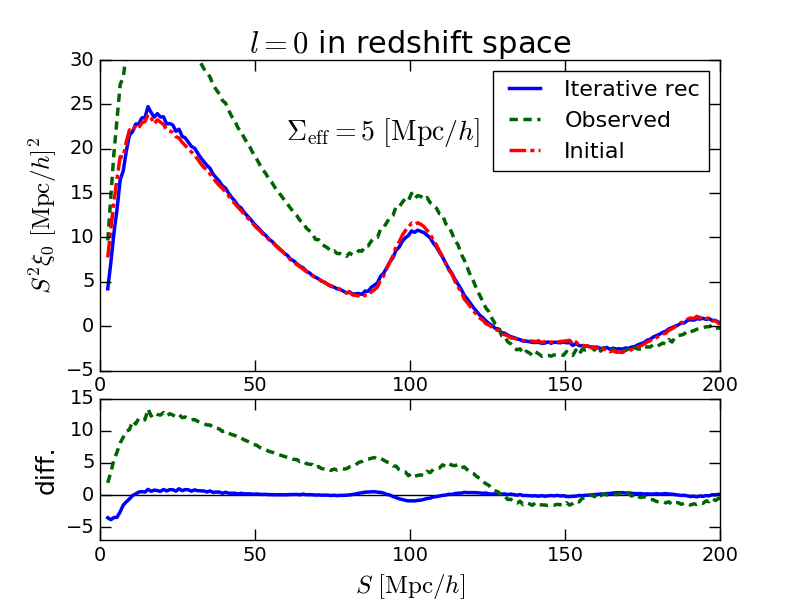}
  \end{center}
\end{minipage}
 \begin{minipage}{0.33\hsize}
  \begin{center}
   \includegraphics[width=65mm]{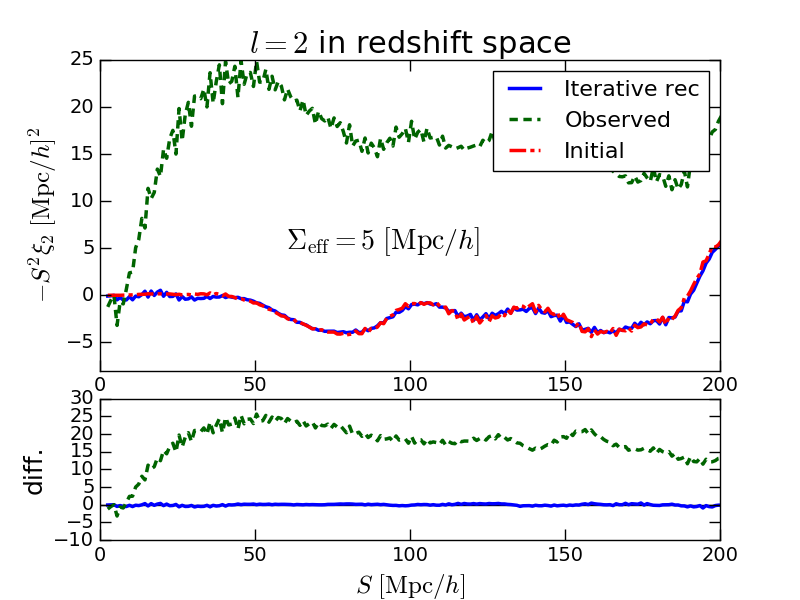}
  \end{center}
\end{minipage} \\
 \begin{minipage}{0.33\hsize}
  \begin{center}
   \includegraphics[width=65mm]{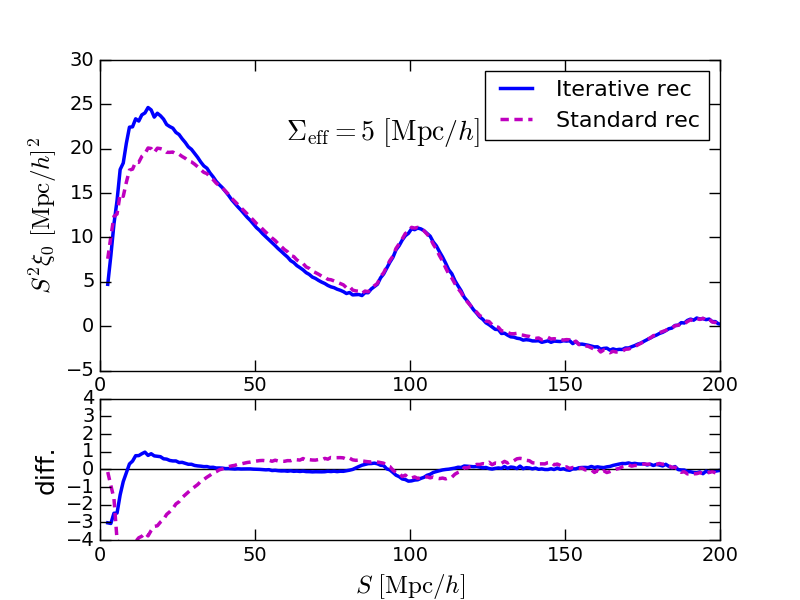}
  \end{center}
\end{minipage}
 \begin{minipage}{0.33\hsize}
  \begin{center}
   \includegraphics[width=65mm]{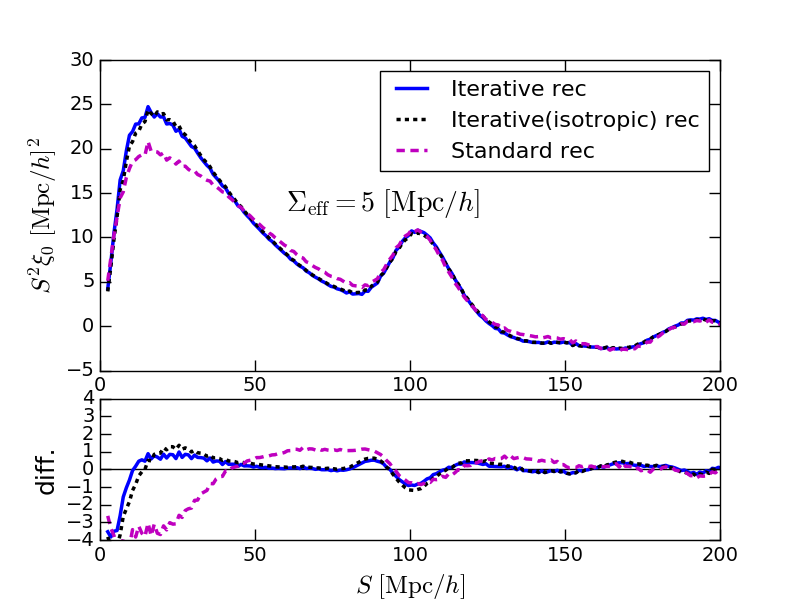}
  \end{center}
\end{minipage}
 \begin{minipage}{0.33\hsize}
  \begin{center}
   \includegraphics[width=65mm]{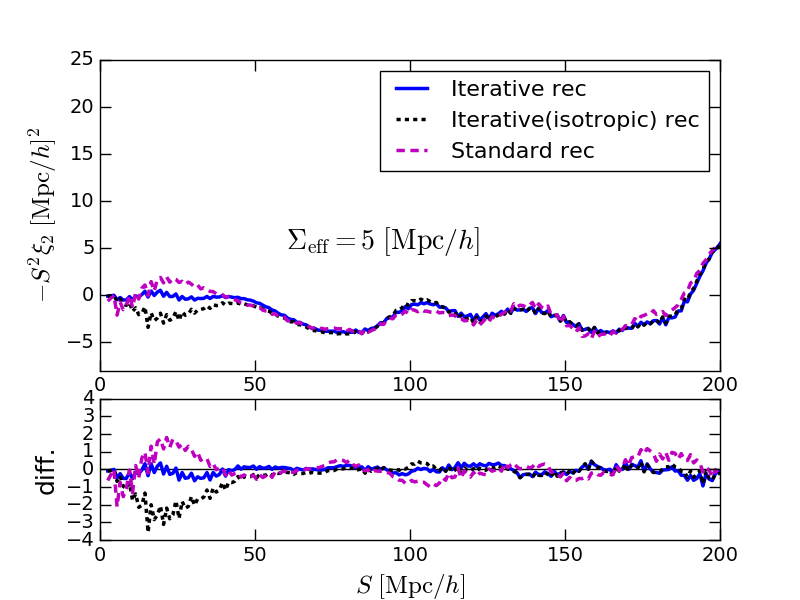}
  \end{center}
\end{minipage}
\end{tabular}
  \caption{\label{2-corr} Two-point correlation function $\xi_{l}(S)$ at $z=0.5$. {\it Left column}: monopole ($l=0$) in real space,  {\it Middle column}: monopole ($l=0$) in redshift space, and {\it Right column}: quadrupole ($l=2$) in redshift space. In the upper row, the solid (in blue), dashed (in green), and dot-dashed (in red) lines correspond to the density field that is reconstructed using our iterative method for $5\hMpc$ smoothing scale, the observed density field, and the initial density field in real space multiplied by the linear growth factor, respectively. The lower row shows the comparison between our iterative method (blue solid) and the standard method for $5\hMpc$ smoothing scale (magenta dashed) (in redshift space, in addition, our iterative method with the isotropic smoothing filtering ($\mathcal{C}_{\rm ani} = 1.0$) (black dotted). The bottom panel in each figure demonstrates the difference with the correlation function for the initial density field.}
\end{figure*}

\subsection{\label{sec:ani_RSD}Effect of the anisotropy in redshift space}

We next consider how the effects of small-scale motions, particularly
the Finger of God effect, impacts on the reconstructed density field.  Fig.~\ref{x_corr_s} shows the cross-correlation coefficient defined as,
\begin{eqnarray}
	s(k) \equiv \frac{\langle \tilde{\delta}_{\rm L} \cdot  \tilde{\delta}^{*}_{\rm ini} \rangle}{ \langle |\tilde{\delta}_{\rm ini} |^2 \rangle}, \label{eq:coe_s}
\end{eqnarray}
where $\tilde{\delta}_{\rm L} $ is the linear density at $z = 0.5$ that is reconstructed with our iterative method using $5\hMpc$ 
smoothing scale in Fourier space and $\tilde{\delta}_{\rm ini}$ is the initial density in Fourier space, multiplied by the linear growth factor so as to extrapolate to $z=0.5$. 

Fig.~\ref{x_corr_s}-(a) corresponds to the coefficient $s(k)$ of the density field that is reconstructed with an isotropic smoothing filter. Ideally the reconstructed density field should be isotropic, with the correlation coefficients independent of the direction of the wavevector; however, we find that there is an inconsistency between the line of sight and the perpendicular directions on smaller scales $k \gsim 0.1 \hMpck$. Based on problem \ref{prob_3} in Section~\ref{sec:Sta}, we can expect that this disagreement is caused by the use of the ``isotropic" smoothing filtering in ``anisotropic" redshift space.

Fig.~\ref{x_corr_s}-(b), (c), and (d) show the dependence of the cross-correlation coefficient $s(k)$ on $\mathcal{C}_{\rm ani}$.  We can see that as $\mathcal{C}_{\rm ani}$ gets larger, the coefficient for the line of sight on intermediate scales, $k\sim 0.2 \hMpck$, increases. Comparing all figures, we can find that in Fig.~\ref{x_corr_s}-(c) ($\mathcal{C}_{\rm ani} = 1.6$), the correlation coefficients for all wavevector directions agree well with each other. In the following analysis in redshift space, we use the value $\mathcal{C}_{\rm ani} = 1.6$.

\subsection{\label{sec:two-corr}Two-point correlation function}

Keeping the above discussion in mind, we turn our attention to the two-point correlation function. In this paper, we particularly focus on its multiple moment defined as,
\begin{eqnarray} 
\xi(S,\mu) = \sum^{\infty}_{l=0}\xi_{l}(S) P_{l}(\mu),
\end{eqnarray}
where $\xi(S,\mu)$ is the two-point correlation function for the density contrast and $P_{l}$ is the Legendre polynomial of order $\ell$. 

The left column of Fig.~\ref{2-corr} shows the monopole in real space and from the upper figure, we can see that our iterative method (blue solid) successfully reconstructs the initial density field (red dot-dashed) on all scales because both lines are almost completely overlapped. Note that the initial correlation function here refers to the actual noisy realization in the simulation, not the ensemble averaged one. This means that our method is recovering the actual supplied density field. Furthermore, the lower figure, where the comparison between our method and the standard method is described, shows that our iterative method considerably improve the standard reconstruction (magenta dashed) on small scales $S \lsim 40 \hMpc$ although there is no difference between the both around on the BAO peak. 
 
The monopole in redshift space is described in the same manner as real space in the middle column of Fig.~\ref{2-corr}. The upper figure shows that also in redshift space, we are able to obtain the initial density field in real space from the final evolved density field with our iterative method. Comparing with the standard method in the lower figure, we find that the reconstruction is improved especially on small scales as it was in real space. In addition, the result of our method using the isotropic smoothing filtering ($\mathcal{C}_{\rm ani} = 1.0$) is showed in black dotted line in the same figure, and the difference with the correlation for the initial condition becomes slightly larger than the one using the anisotropic smoothing filtering ($\mathcal{C}_{\rm ani} = 1.6$) on small scales. 

In the right column of Fig.~\ref{2-corr}, we show the quadrupole in redshift space. We can apparently find that our method, taking account of the anisotropy discussed in Section~\ref{sec:ani_RSD}, is able to correctly handle with the effect of the redshift-space distortion (from the upper figure) and that the performance is better than the ones using the isotropic smoothing filtering and the standard method (from the lower figure).

\section{Conclusions} \label{sec:Conc}
 
Based on \citet{1999MNRAS.308..763M}, we have presented an iterative method to reconstruct the initial linear density field. To evaluate the performance of this method, first, we showed the cross-correlation between the reconstructed displacement field and the true displacement field of the dark matter particles  We found that our method can estimate the displacement up to smaller scales compared with the standard method both in real and redshift space. Furthermore, we investigated the effects of the anisotropy in redshift space on the reconstruction process by comparing the cross-correlation functions between the reconstructed density field and the initial density field that are averaged over various directions.  We introduced the following parameter to manage the effects: $\mathcal{C}_{\rm ani}$, which is corresponding to the ratio of the smoothing scale for the line of sight direction to the perpendicular direction. Finally, we found that our iterative method is able to restore the two-point correlation function more successfully than the standard method especially on small scales.

We comment on the limitations that the previous reconstruction methods have (see Section~\ref{sec:Sta} and \ref{sec:discu}). We found that our iterative method converges successfully in a practical case, which means that it can find an initial density field that will evolve into the present galaxy density field given our ansatz for the separation of large-scale and small-scale modes.  In addition, we considered only the first-order perturbation theory when estimating the displacement, assuming that we will apply it to realistic galaxy density fields. On this point, we need to evaluate in a future paper how the results are influenced by galaxy bias by applying our method to the halo-based mock galaxy catalogs.

Furthermore, regarding the redshift-space distortion modeling, we found that the anisotropic smoothing filtering actually improves the two-point correlation on small scales (see Section~\ref{sec:two-corr}). However, we  emphasize that the initial correlation on intermediate and large scales is reconstructed well by our method even when we use the isotropic smoothing filtering.  This fact suggests that our iterative method could be applied to the measurement of the linear growth rate $f$.

In our method, once we distribute the galaxy particles on a grid, we subsequently perform the iterative process on the grid without using the original particles. Therefore, we are not bothered with the problem moving the galaxies and the random particles differently in redshift space. In addition, reconstruction methods on the basis of grid have the advantage that the computational time is determined mainly by the number of grid cells even if the number of galaxies obtained from future galaxy surveys gets larger. 

Finally, our investigations here has restricted our analysis to the plane-parallel approximation, in which we treat the line of sight as a fixed direction (i.e., $\hat{\bf z}$ in Eq.~(\ref{eq:s})). This approximation is broken in realistic wide-field surveys \citep[e.g.][for the "bull's-eye effect"]{1994ApJ...422...46P,1997ApJ...479L..15P,2004ApJ...601...28T}, and thus we will need to extend the method. This is straight-forward: in our iterative method, we are constructing the displacement field in configuration space, so that we can easily isolate the velocity along the line of sight when applying the redshift-space distortion to the displacement field.   

In upcoming work, we will extend our method to biased galaxy samples toward easy application to the next-generation galaxy redshift surveys. Furthermore, we intend to shorten the computation time because we need to apply the reconstrucntion method to a large number of realizatioins to estimate the sample covariance matrices that we use in the fitting.

\section*{Acknowledgements}

We would like to thank M. Takada, T. Sunayama, and T. Nishimichi for useful discussions. RH is supported by Japan Society for the
Promotion of Science, Research Fellowships for Young Scientists (No. 16J01773) and as a doctoral course student in Division for
Interdisciplinary Advanced Research and Education, Tohoku University. DJE is supported by U.S.\ Department of Energy grant DE-SC0013718 and as a Simons Foundation Investigator.

\bibliographystyle{mnras}
\bibliography{ms}

\bsp	
\label{lastpage}
\end{document}